\newif\ifreview

\reviewfalse

\ifreview
\documentclass[a4paper,12pt,review]{elsarticle}
\usepackage{geometry}
\usepackage{amsmath,amssymb}
\biboptions{authoryear}
\newcommand{\shorttitle}[1]{}
\newcommand{\shortauthors}[1]{}
\newcommand{\titlewrap}[1]{\title{#1}}
\newcommand{\authorwrap}[3]{\author[#1]{#2}}
\newcommand{\cormarkwrap}[1]{}
\newcommand{\credit}[1]{}
\newcommand{\bKW}{\begin{keyword}}
\newcommand{\eKW}{\end{keyword}}
\makeatletter 
\def\ps@pprintTitle{%
 \let\@oddhead\@empty
 \let\@evenhead\@empty
 \def\@oddfoot{}%
 \let\@evenfoot\@oddfoot}
\makeatother 
\newcommand{\printcredits}{}
\usepackage{setspace,caption}
\newcommand{\figwthS}{0.7\textwidth}
\newcommand{\figwthL}{\textwidth}

\else
\documentclass[a4paper,fleqn]{cas-dc}
\usepackage[authoryear]{natbib}
\newcommand{\bKW}{\begin{keywords}}
\newcommand{\eKW}{\end{keywords}}
\newcommand{\figwthS}{\columnwidth}
\newcommand{\figwthL}{\textwidth}
\newcommand{\titlewrap}[1]{\title[mode = title]{#1}}
\newcommand{\authorwrap}[3]{\author[#1]{#2}[#3]}
\newcommand{\cormarkwrap}[1]{\cormark[#1]}
\fi

\usepackage[normalem]{ulem}

\usepackage[switch]{lineno}
\ifreview
\linenumbers

\let\oldequation\equation
\let\oldendequation\endequation

\renewenvironment{equation}
  {\linenomathNonumbers\oldequation}
  {\oldendequation\endlinenomath}
  
\fi  

\newcommand{\nuclen}{h_\mathrm{n}}
\newcommand{\corlen}{\xi_0}

\newcommand{\patchsize}{a_\mathrm{max}}
\newcommand{\taupmin}{\tau_\mathrm{p}^\mathrm{min}}
\newcommand{\taupmax}{\tau_\mathrm{p}^\mathrm{max}}
\newcommand{\tauamp}{\tau_\mathrm{amp}}
\newcommand{\taup}{\tau_\mathrm{p}}

\newcommand{\TC}{T_\mathrm{C}}
\newcommand{\taucr}{\tau_\mathrm{cr}}
\newcommand{\taucrA}{\tau_\mathrm{cr}^\mathrm{A}}
\newcommand{\taucrB}{\tau_\mathrm{cr}^\mathrm{B}}
\newcommand{\taucrC}{\tau_\mathrm{cr}^\mathrm{C}}
\newcommand{\taun}{\tau_\mathrm{n}}
\newcommand{\taur}{\tau_\mathrm{r}}
\newcommand{\tauf}{\tau_\mathrm{f}}
\newcommand{\taukin}{\tau_\mathrm{kin}}
\newcommand{\tausin}{\tau_{\sin}}

\begin{document}
\let\WriteBookmarks\relax
\def\floatpagepagefraction{1}
\def\textpagefraction{.001}
\shorttitle{Friction nucleation by microslip coalescence}
\shortauthors{S. Sch\"ar et~al.}

\titlewrap{Nucleation of frictional sliding by coalescence of microslip} 


\authorwrap{1}{Styfen Sch\"ar}{}
\credit{Data curation, Writing - Original draft preparation}

\authorwrap{1,2}{Gabriele Albertini}{orcid=0000-0001-9565-7571}
\credit{Data curation, Software, Writing - Original draft preparation}

\authorwrap{1}{David S. Kammer}{orcid=0000-0003-3782-9368}
\cormarkwrap{1}
\ead{dkammer@ethz.ch}
\credit{Conceptualization, Methodology, Software, Writing - Review \& Editing}

\address[1]{Institute for Building Materials, ETH Zurich, Switzerland}
\address[2]{School of Civil and Environmental Engineering, Cornell University, Ithaca, NY, USA}

\cortext[cor1]{Corresponding author}

\begin{abstract}
The onset of frictional motion is mediated by the dynamic propagation of a rupture front, analogous to a shear crack. The rupture front nucleates quasi-statically in a localized region of the frictional interface and slowly increases in size. When it reaches a critical nucleation length it becomes unstable,  propagates dynamically and eventually breaks the entire interface, leading to macroscopic sliding. The nucleation process is particularly important because it determines the stress level at which the frictional interface fails, and therefore, the macroscopic friction strength.  However, the mechanisms governing nucleation of frictional rupture fronts are still not well understood. Specifically, our knowledge of the nucleation process along a heterogeneous interface remains incomplete. Here, we study the nucleation of localized slip patches on linear slip-weakening interfaces with deterministic and stochastic heterogeneous friction properties. Using numerical simulations, we analyze the process leading to a slip patch of critical size for systems with varying correlation lengths of the local friction strength. Our deterministic interface model reveals that the growth of the critical nucleation patch at interfaces with small correlation lengths is non smooth due to the coalescence of neighboring slip patches. Existing analytical solutions do not account for this effect, which leads to an overestimation of global interface strength. Conversely, when the correlation length is large, the growth of the slip patch is continuous and our simulations match the analytical solution. Furthermore, nucleation by coalescence is also observed on stochastic interfaces with small correlation length. In this case, the applied load for a given slip patch size is a random variable. We show that its expectation follows a logistic function, which allows us to predict the strength of the interface well before failure occurs. Our model and observations provide new understanding of the nucleation process and its effect on the static frictional strength.
\end{abstract}


\bKW
friction \sep nucleation by coalescence \sep random interface
\eKW

\maketitle

\section{Introduction}
\label{sec:introduction}


Friction occurs in many mechanical, biological and geophysical systems. In some systems, it is undesirable because it decreases efficiency and causes wear. While in other systems, friction is beneficial because it provides a stabilizing force, \textit{e.g.}, walking would be impossible without friction.  
Due to its relevance, friction has been extensively studied over the last centuries. However, a complete fundamental understanding, especially concerning the nucleation of frictional sliding, is still missing. 

The oldest and most widely known friction model is the Amontons-Coulomb friction law, which states that the friction strength is proportional to the applied normal force with the proportionality constant being the friction coefficient $\mu_s$
\citep{amontons_resistance_1699,coulomb_theorie_1785,popova_research_2015}. 
The reason for this proportionality was explained by \cite{bowden_friction_1950}. 
They observed that the real contact area of contacting rough surfaces is considerably smaller than the apparent area, implying that the normal stress at the contact points reaches the material hardness $\sigma_\mathrm{H}$. Thus, to maintain equilibrium, the real area of contact is proportional to the applied normal stress \citep{dieterich_imaging_1996}. Assuming that the frictional strength is the sum of the shear strength of the micro-contacts, $\tau_s$, this results in the friction coefficient being the shear-strength-to-hardness ratio, $\mu_s=\tau_s/\sigma_{\mathrm{H}}$, of the micro-contacts \citep{}.

Many studies have been dedicated to refining this friction law by including rate dependency and aging. Direct observation of the evolution of real contact area revealed that it increases with contact age \citep{dieterich_direct_1994,dillavou_nonmonotonic_2018} and decreases with changes in shear stress well before the onset of frictional sliding \citep{sahli_evolution_2018,dillavou_shear_2020}. 
These state-dependent properties of friction can be taken into account by state variables with a given evolution law. In addition, slip-rate dependent properties have been reported on multiple materials such as rock \citep{kilgore_velocity_1993}, metals \citep{rabinowicz_intrinsic_1958} and polymer glasses \citep{baumberger_solid_2006}.
A widely used family of models combines these effects into rate-and-state friction laws \citep{ruina_slip_1983,marone_laboratory-derived_1998,selvadurai_modeling_2020}.

Interestingly, it is common to see large variations of friction coefficients in experiments with seemingly identical setups \citep{rabinowicz_friction_1992,ben-david_static_2011}, even though friction laws are generally deterministic. The reason is that most friction laws are based on single degree-of-freedom considerations assuming the interaction between two undeformable blocks. This is a very limiting assumption and does generally not hold for many systems that are relatively large and compliant. These systems typically show stable localized sliding, while large parts of the interface are still stuck \citep{nielsen_experimental_2010,latour_characterization_2013,mclaskey_foreshocks_2013}. Eventually, the sliding region becomes unstable, propagates dynamically -- like a crack -- and, once it ruptures the entire frictional interfaces, leads to macroscopic sliding \citep{svetlizky_classical_2014,kammer_linear_2015,rubino_understanding_2017,svetlizky_brittle_2019,svetlizky_dynamic_2020}.

Although, the presence of large variation in frictional strength is well known and the nucleation of frictional sliding has been studied theoretically \citep{uenishi_universal_2003,uenishi_three-dimensional_2018,rubin_earthquake_2005, ampuero_properties_2006,ampuero_earthquake_2008,ray_earthquake_2017,de_geus_how_2019} as well as experimentally \citep{nielsen_experimental_2010,latour_characterization_2013,mclaskey_foreshocks_2013}, the link between local variations in interface properties and the observed variations in macroscopic strength remains poorly understood.

Recently, \cite{albertini_stochastic_2020} demonstrated, using numerical simulations, that the macroscopic strength of a random interface can be quantitatively well predicted with an analytical solution \citep{uenishi_universal_2003,ampuero_properties_2006}, which determines a critical nucleation length and its associated stress level. The simulations confirmed that macroscopic sliding occurs when a slipping region reaches this critical length. They further showed that random interfaces with decreasing correlation length of the local strength present increasing global strength. However, the simulations also suggested that there is an increasing discrepancy between the theoretical prediction and the observations in the simulations for decreasing correlation length. The cause of this discrepancy and its effect have not been explored, but are important for realistic systems because asperities, one of the main origins of randomness at interfaces, are usually much smaller than the nucleation length \citep{pantcho_stoyanov_scaling_2017} and, therefore, the interface correlation length is expected to be comparably small.


In the present work, we aim to understand the growth process that leads to this critical nucleation length and, therefore, provide fundamental knowledge about the macroscopic strength of random interfaces with small correlation length. We will demonstrate that heterogeneities on a very small length scale result in the same critical nucleation length as for systems with intermediate and large correlation length. However, the growth process leading to nucleation is very different. While large heterogeneities cause continuous growth, as expected from previous work, small heterogeneities lead to growth by coalescence, which results in lower overall strength compared to existing theoretical predictions. 

This paper is organized as follows. First, in Sec.~\ref{sec:materialmethods} we will introduce the physical system used for dynamic simulations, present our approach to generate deterministic frictional interfaces and summarize the method used by \cite{albertini_stochastic_2020} to generate random interfaces with a specific correlation length and probability density. In Sec.~\ref{sec:theory}, we describe an analytical approach to predict the frictional strength of our dynamic simulations. The nucleation process of frictional sliding for different correlation lengths is presented in Sec.~\ref{sec:results}. Further, we show how the accuracy of the analytical approach depends on the correlation length and we verify our findings on a random interface. We then consider the growth of the largest slip patch as a stochastic process and, therefore, predict the macroscopic frictional strength before the onset of global sliding.
Finally, we discuss our results on the nucleation process, the analytical approach and the prediction of the global strength in Sec.~\ref{sec:discussion} and draw a conclusion in Sec.~\ref{sec:conclusion}.

\section{Material and methods}
\label{sec:materialmethods}
In this section, we describe the physical model used and the dynamic simulations performed to investigate the nucleation of frictional sliding. We will study deterministic frictional interfaces, which we will use to precisely analyze the nucleation and growth of slip patches. Further, we will verify the observed formation patterns of critical slip patches on random frictional interfaces with specific stochastic properties.

\begin{figure}
    \centering
    \includegraphics[width=\figwthS]{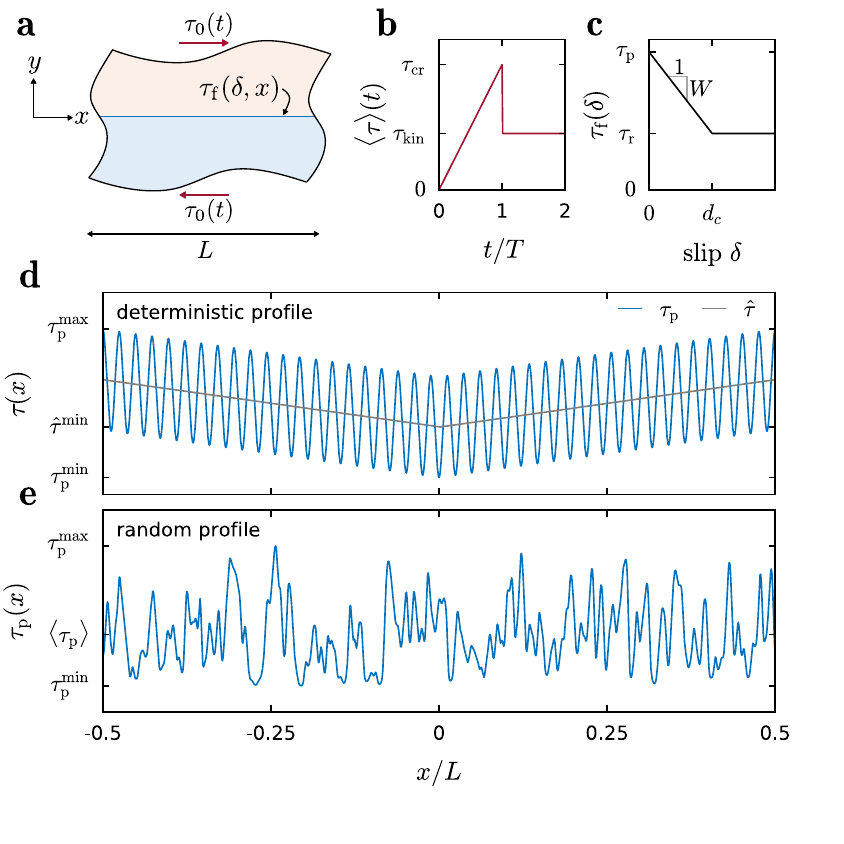}
    \caption{Problem statement. 
    (a)~Two 2D linear elastic solids with a frictional interface subjected to a uniform shear load $\tau_0$. The solid half spaces are periodic in $x$-direction with repetition length $L$, and infinite in $y$-direction. 
    (b)~Average interface stress $\langle \tau \rangle$ increases quasi-statically in time to the critical value $\taucr$ where the interface becomes globally unstable and macroscopic sliding occurs. The onset of sliding results in a drop from the peak strength (static friction) to the kinetic strength $\taukin$ (sliding friction). 
    (c)~Linear slip-weakening law of the interface. The local interface strength $\tauf$ declines linearly with increasing slip at a weakening rate $W$ from its peak $\taup$ to its residual strength level $\taur$. 
    (d)~Deterministic profile with frictional peak strength oscillating around its base value $\hat \tau$. The maxima are located at the boundaries of the profile and the minimum at the center. 
(e)~Random profile with a correlation length $\corlen=0.05\nuclen$ (see definition of the nucleation length $\nuclen$ in Eq.~\ref{eq:hn}).}
    \label{fig:ProblemStatement}
\end{figure}

\subsection{Physical model}\label{sec:matmet:physicalmodel}
We study how microscopic heterogeneities at frictional interfaces influence their macroscopic frictional strength with a focus on the nucleation process.
We consider two semi-infinite solids, as shown in Fig.~\ref{fig:ProblemStatement}a. The solids are infinite in the $y$-direction and periodic in $x$ with repetition length $L$. Because we consider thin plates, a two-dimensional plane-stress system is used. Both half spaces have the same linear elastic material properties with a shear modulus of $G=1$~GPa, a Poisson's ratio of $v=0.33$ and a density of $\rho=1170$~kg/m$^3$. 
These values are related to glassy polymers, which are commonly used in experimental friction and fracture mechanics.

On a macroscopic level the system is loaded with a shear load $\tau_0(t)$. The load is uniformly distributed along the interface and increases quasi-statically at a constant rate (see Fig.~\ref{fig:ProblemStatement}b). When $\tau_0$ exceeds the macroscopic strength $\taucr$ of the interface, it becomes unstable and starts sliding. This global instability results in a stress drop to the kinetic level of friction $\taukin$. The drop in frictional strength is associated with the transition from static friction to sliding friction.

The macroscopic observation is related to the microscopic level, where we consider a linear slip-weakening friction law (Fig.~\ref{fig:ProblemStatement}c),
\begin{equation}\label{eq:slip_weakening}
    {\tauf}(\delta ) = {\taur} + W({d_c} - \delta )H({d_c} - \delta ),
\end{equation}
which states that the interface strength $\tauf$ decreases from its peak $\taup$ to its residual strength $\taur$ at a constant weakening rate $W=(\taup -\taur)/d_c$ over a characteristic slip-length $d_c$. Note that $H(.)$ is the Heaviside function.
When the local shear stress level reaches the local peak strength $\taup$ within the interface, it initiates the onset of slip $\delta$. Slip is accompanied by a reduction of the frictional strength $\tauf(\delta)$ according to the slip-weakening law in Eq.~\ref{eq:slip_weakening}. Because of the residual strength, we observe a plateau of the frictional strength at $\taukin$ on the macroscopic level. In this work, without loss of generality, we set $\taur$ to zero. 
Since, we consider a uni-material interface, the normal stress $\sigma$ remains constant during the nucleation and propagation phase of the friction onset.
Thus, a possible coupling of local normal stress  to local friction strength similar to the Amontons-Coulomb law would yield the same results.
The assumption of using a linear slip-weakening friction law is discussed in Sec.~\ref{sec:discussion}.

\subsection{Dynamic simulations} \label{sec:matmet:simulation}
We solve the physical model described in Sec.~\ref{sec:matmet:physicalmodel} with numerical simulations. Since our interfaces are periodic in $x$, we solved the elasto-dynamic equations efficiently with the spectral boundary integral method \citep{geubelle_spectral_1995,geubelle_numerical_1997}. Due to symmetry, only one half-space was solved and we verified convergence with respect to spatial discretization, loading rate and time step.

\subsection{Deterministic frictional interface}\label{sec:matmet:deterministic}
In a first step, we will study the microscopic nucleation process of frictional sliding in a system with deterministic distribution of the local peak strength $\taup(x)$ along the interface (see Fig.~\ref{fig:ProblemStatement}d). This will allow us to gain a fundamental understanding of the process, which we then use in a second step for interfaces with random strength profiles, as outlined in Sec.~\ref{sec:matmet:stochastic}. The deterministic strength profile $\taup$ is a superposition of a triangular base strength $\hat \tau$ and a sinusoidal component $\tausin$ with the period $P$ and amplitude $\tauamp$:
\begin{equation}\label{eq:det_interface}
\begin{aligned}
  &\taup(x) &&= \hat \tau (x) - \tausin(x), \\
  &\hat \tau (x) &&= m\left| {x} \right| + {{\hat \tau }^{\min }}, \\
  &{\tau _{\sin }}(x) &&= {\tau _{amp}}\cos \left( {2\pi x/P} \right). 
\end{aligned}
\end{equation}
The slope $m$ is fixed to $10$~MPa/m and $\hat \tau^{min}$ to $0.9$~MPa. 
Consequently, the initial slip nucleation is forced to the center of the profile. 
We define $L$ as an odd multiple of the period $P$ to ensure a smooth continuity with $\taupmax$ across the repetition length of the system.
To separate the length scale $L$ from the one of the critical nucleation length $\nuclen$ (see Sec.~\ref{sec:theory}), $L$ is set to at least four times $\nuclen$. We verified on selected simulations that a larger $L$ does not affect the results.

\subsection{Stochastic frictional interface}\label{sec:matmet:stochastic}
In addition to the deterministic interface profile, we also study the nucleation of frictional sliding on random interfaces. We generate profiles of random local peak strength $\tau_p(x)$ that exhibit specific stochastic properties (see Fig.~\ref{fig:ProblemStatement}e). 
We aim to generate random frictional interfaces with given spectral density function and probability distribution, and, therefore, adopted the following procedure \cite[ch. 5]{grigoriu_stochastic_2013}, which was also used by \cite{albertini_stochastic_2020}.

A Gaussian random field $Z(x)$ with two sets of independent Gaussian random variables $A_j$ and $B_j$ with zero mean and unit variance was constructed following
\begin{equation}\label{eq:tauc2}
    {Z}(x) = \sum\limits_{j = 1}^J {{\sigma _j}\left( {{A_j}\cos ({k_j}x) + {B_j}\sin ({k_j}x)} \right)},
\end{equation}
and $\sigma_j$ was normalized to ensure that $Z(x)$ has unit variance
\begin{equation}
    \sigma _j^2 = \frac{{g({k_j})}}{{\sum\nolimits_{j = 1}^J {g({k_j})} }}.
\end{equation}
The wave numbers $k_j=k2\pi/L$ make the profile $L$-periodic. The spectral density function, $g(k)$, is defined as the Fourier transform of the correlation function of $Z$, $C_z(.)$, and is given by
\begin{equation}
    g(k) = \int_{ - \infty }^\infty  {{C_Z}(\xi ){e^{-ik\xi }}\mathrm{d}\xi } ,
\end{equation}
which decays with a power law $g(k) \propto {({k^2} + {\lambda ^2})^{ - 4}}$. 
The correlation length $\corlen$ is inversely proportional to $\lambda$: $\corlen=2\pi/\lambda$. The Gaussian random field was transformed into a field following a Beta distribution by applying the cumulative density function (CDF) of a standard normal distribution $\phi(.)$ and the inverse CDF of a Beta distribution $F^{-1}(.)$ with the parameters $\alpha=1.5$ and $\beta=3$
\begin{equation}\label{eq:tauc1}
    {\tau _p(x)} = {\taupmin} + \left({\taupmax} - {\taupmin}\right){F^{ - 1}}\Big({\phi \big(Z(x)\big)}\Big).
\end{equation}
$\taupmin$ and $\taupmax$ were set to $0.667$ MPa and $1.667$ MPa, respectively. These values result in a mean value for $\tau_p$ of $1$ MPa with a standard deviation of $0.2$ MPa.
The nonlinear mapping $F^{-1}\circ \phi$ in Eq.~(\ref{eq:tauc1}) only has a minor effect on the correlation function, such that $C_Z(\xi)\approx C_{\taup}(\xi)$. The approximate conservation of the correlation function after applying a nonlinear mapping is a property of positive correlation functions $C_Z(\xi)>0$ \citep[p.48]{grigoriu_applied_1995}.

\section{Theory}
\label{sec:theory}

In a frictional system, slip initiates when the loading stress locally exceeds the strength level of the interface. Slip goes along with a local reduction in stress level due to a slip-weakening process (Eq.~\ref{eq:slip_weakening}). Consequently, for the system to remain stable, the load is partially redistributed on regions with a stress level below their strength. With increasing load, more and more stress is redistributed from slipping areas to ever shrinking sticking parts. Eventually, the system cannot find any static solution to carry the load anymore, and it fails. This leads to a dynamic propagation of a slip front until the entire interface is sliding. 

\cite{uenishi_universal_2003} studied nucleation on  slip-weakening interfaces with uniform friction properties and a locally peaked load. 
They showed that the stability of a slip patch can be reduced to an eigenvalue problem, and that the critical nucleation length $\nuclen$, which leads to instability, is universal.
In particular, they showed that $\nuclen$ is independent of the spatial distribution of the load and depends only on the material and friction properties:
\begin{equation}\label{eq:hn}
    {\nuclen} \approx 1.158\frac{{{\mu ^*}}}{W} ~,
\end{equation}
where $\mu^*$ is an effective shear modulus. \cite{ampuero_properties_2006} extended this approach from a system with a locally peaked load to a random stress distribution. 

Recently, \cite{albertini_stochastic_2020} reformulated the problem for an interface with a random strength profile and defined a nucleation stress $\taun(x)$, which is given by
\begin{equation}\label{eq:taun}
{\taun}(x) \approx 0.751\int_{-1}^{1} {\taup\left[ {({\nuclen}/2)s + x} \right]{v_0}} (s)\mathrm{d}s ~,
\end{equation}
where ${\nu _0}(s) \approx (0.925 - 0.308{s^2})\sqrt {1 - {s^2}}$ is the first eigenfunction of the elastic problem. Note that this assumes $W$ to be a constant. 
$\taun(x)$ is the stress required to nucleate a slip patch of size $\nuclen$ centered at $x$.

The critical load $\taucr$ for the entire interface is the stress at which the first slip patch reaches a nucleation length of size $\nuclen$. Hence, it corresponds to the minimum of $\taun(x)$:
\begin{equation}\label{eq:taucr_theory}
    \taucr=\min(\taun(x)) ~.
\end{equation}
Note that in all simulations presented here, we set $d_c(x)$ such that $W$ is uniform (see Eq.~\ref{eq:slip_weakening}). Therefore, Eq.~\ref{eq:taun} is applicable and the nucleation length is expected to be constant.

\section{Results}
\label{sec:results}

We will first present simulation results for the deterministic case in Sec.~\ref{sec:results:deterministic} and compare our observations with theoretical predictions from Sec.~\ref{sec:theory}. We will identify the range of parameters for which the theoretical solution fails to predict accurately the actual interface strength and show that the discrepancy is caused by a different growth mechanism based on coalescence of localized slip patches. In the second part, in Sec.~\ref{sec:results:stochastic}, we will present simulation results for the stochastic case, which exhibit the same coalescence process, and analyze the growth rate of the slip patches.

\subsection{Deterministic interfaces}
\label{sec:results:deterministic}

\subsubsection{Nucleation patterns}\label{subsubsec:NucleationPattern}

\begin{figure*}
    \centering
    \includegraphics[width=\figwthL]{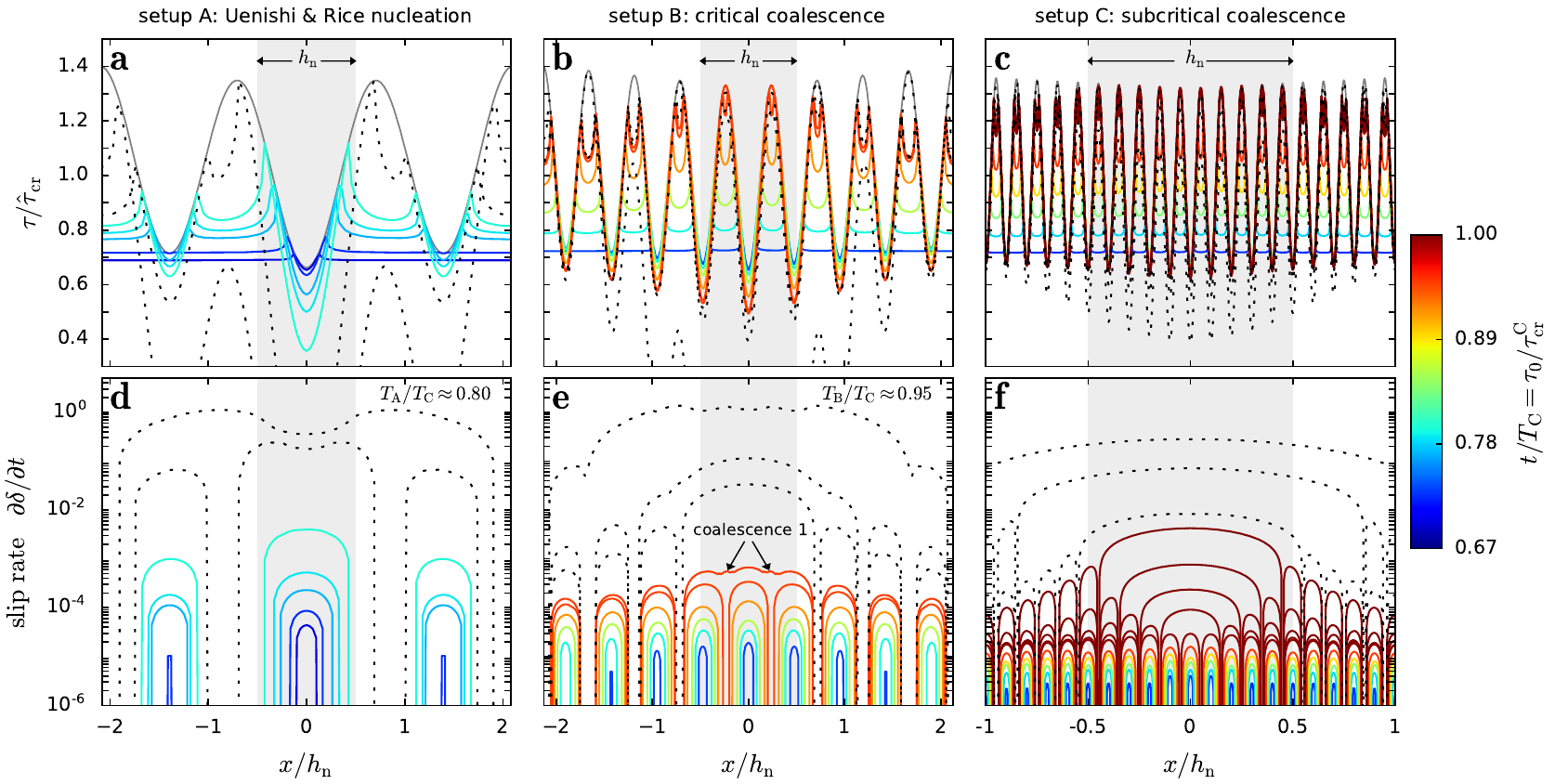}
    \caption{Nucleation of slip in deterministic profiles. Setup A corresponds to a period $P=1.4$ $\nuclen$, setup B to $P=0.48$ $\nuclen$ and setup C to $P=0.1$ $\nuclen$. (a), (b) and (c) show the profile as a gray line and the stress levels at various times as colored lines. The times $t$ are normalized by the time $\TC$ at which setup C fails. For each setup, the stress states plotted in color correspond to stable systems and the dashed lines show unstable stress states after failure. $\tau(x)$ is normalized by $\hat \taucr$, which is the theoretical strength of the deterministic profile $\taup(x) = \hat \tau(x)$ (Eq.~\ref{eq:det_interface}) without the sinusoidal component. (d), (e) and (f) show the slip rate and, therefore, the extension of the slip area within the profile at the same time steps as the stresses in (a), (b), and (c), respectively. (c) and (f) show, for visual purposes, only a subsection of the simulated domain, \textit{i.e.}, $L/\nuclen \geq 4$.}
    \label{fig:DeterministicInterface}
\end{figure*}

We run simulations of interfaces with deterministic profiles of varying characteristic length scale $P$, ranging between $P/\nuclen = 0.1 - 1.0$, and we study the  effect of $P$ on the growth process leading to nucleation. Three representative cases are shown in Fig.~\ref{fig:DeterministicInterface}. In all simulations, the load is slowly increased over time $\tau_0=Rt$. When the local shear stress $\tau(x)$ reaches $\taup(x)$ at any point within the interface, slip initiates locally and the stress level starts to decrease due to the slip-weakening process (Eq.~\ref{eq:slip_weakening}). At time $t/\TC \approx 0.7$, for instance, all three cases shown in Fig.~\ref{fig:DeterministicInterface} present localized slip patches (see blue curves), which are co-located at the minima of the $\tau_p(x)$ profile. In setup A, there are three slip patches located at $x/\nuclen \approx 0,$ and $\pm1.4$. Setup B has $7$ slip patches at that time and setup C many more. These early stage slip patches are stable and the system remains in equilibrium.

\begin{figure}
    \centering
    \includegraphics[width=\figwthS]{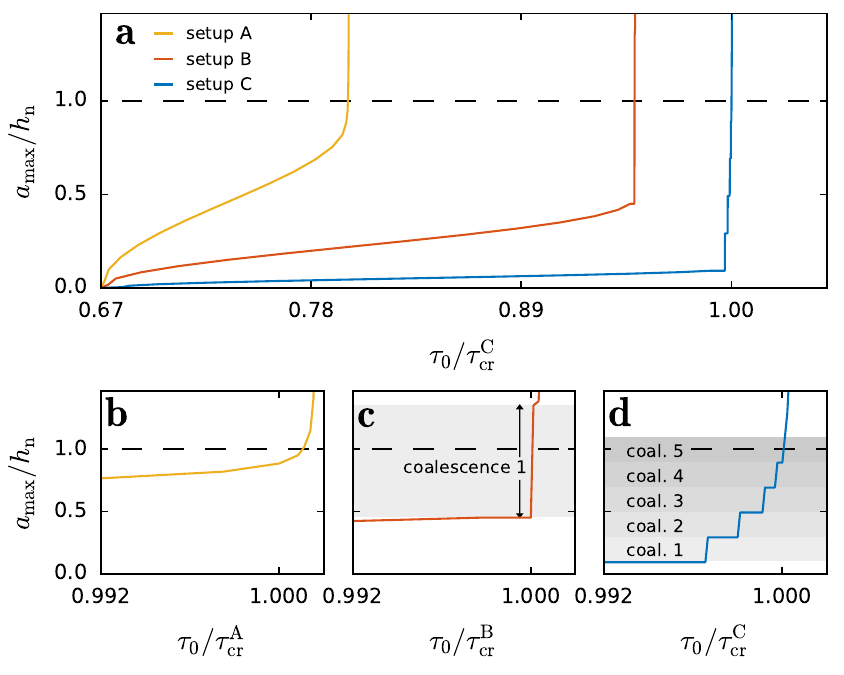}
    \caption{Growth of the critical slip patch. (a) Size of the critical slip patch of setup A, B and C. Stable slip initiates for all profiles at $\tau_0=0.67\taucrC=\taupmin$. (b) Final phase of crack growth of setup A: Continuous patch growth until the profile fails at $\tau_0=\taucrA$. (c) Final growth phase of the critical nucleation patch of setup B. At $\tau_0=\taucrB$ the central patch coalesces with its neighboring patches. The size of the critical patch rises sharply by coalescence and initiates a dynamic slip front.
    (d) The final growth phase of the critical patch of setup C is governed by multiple coalescence events of the central patch with its adjacent patches. The fifth coalescence results in dynamic failure at $\tau_0 = \taucrC$.}
    \label{fig:Patchgrowth}
\end{figure}

As the load continuous to increase, the slip patches grow in size and local slip rate increases. In setup A, this slow and stable growth of the slip patches continues until the most critical one, which is located at $x/\nuclen = 0$ reaches the critical length $\nuclen$ (highlighted by gray area) at $t/\TC \approx 0.80$ and becomes unstable. At this moment, it propagates dynamically through the interface and causes global sliding. In this process, it coalesces with the other slip patches. However, this is a post critical mechanism. The smooth growth process of the critical slip patches is also shown in Fig.~\ref{fig:Patchgrowth}a and b. Note the change in slope in Fig.~\ref{fig:Patchgrowth}b when $\patchsize$ reaches $\nuclen$, which highlights the transition towards unstable growth. This process corresponds to the assumption of the theoretical solution presented in Sec.~\ref{sec:theory}, which is why we name this here a Uenishi \& Rice nucleation.

In setup B, the nucleation process is different. First, all slip patches grow stably, as in setup A, but at $t/\TC \approx 0.95$, the central slip patch coalesces with its two neighbors (see Fig.~\ref{fig:DeterministicInterface}e). This causes a jump in the size of the largest slip patch, as shown in Fig.~\ref{fig:Patchgrowth}a and c. This happens because the characteristic length of the $\taup(x)$ profile is considerably smaller than the critical nucleation length, \textit{i.e.}, $P/\nuclen = 0.48$. Interestingly, the coalescence leads directly to a slip patch that is larger than the critical nucleation length (see Fig.~\ref{fig:DeterministicInterface}e and Fig.~\ref{fig:Patchgrowth}c), and hence the interface becomes unstable and transitions to global sliding. During this process, the unstable slip front dynamically coalesces with the other stable patches. 
In this setup B, nucleation occurred with the first slip-patch coalescence, when the largest stable slip patch was far from its critical length, \textit{i.e.}, $\patchsize/\nuclen < 0.5$ before coalescence. 
We name this mechanisms a nucleation by critical coalescence. 

Finally, the nucleation processes in setup C and setup B are similar. However, in setup C, $P/\nuclen \ll 1$, which results in multiple coalescence events before failure (see Fig.~\ref{fig:DeterministicInterface}f and Fig.~\ref{fig:Patchgrowth}d). Due to the small characteristic interface length $P$, the first $4$ coalescence events do not lead to $\patchsize/\nuclen > 1$ (see Fig.~\ref{fig:Patchgrowth}d). Hence, stable growth continues until eventually a coalescence event leads to a super-critical slip patch and instability. We call this process a nucleation by sub-critical coalescence.

\subsubsection{Comparison with theoretical solution}
\label{sec:results:prediction}

\begin{figure}
    \centering
    \includegraphics[width=\figwthS]{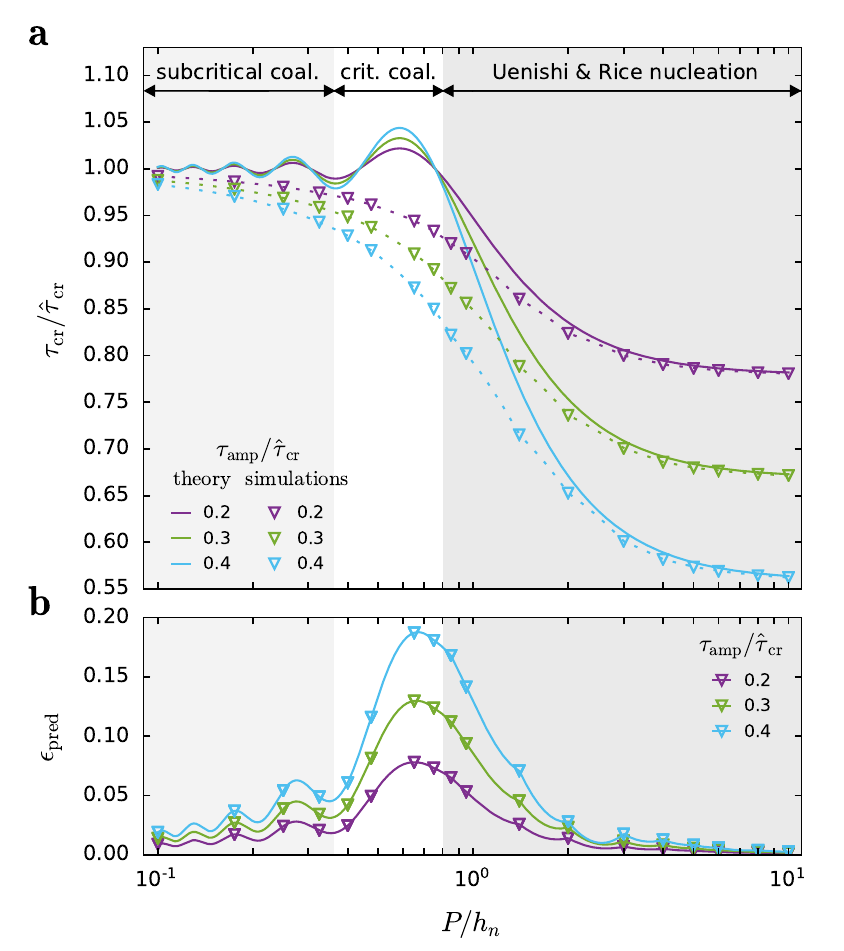}
    \caption{Comparison of the simulation results with the analytical solution. (a) The dashed line is a linear interpolation between the simulation results depicted as triangles. The continuous lines show the analytical solution with Eq.~\ref{eq:det_interface} applied in Eq.~\ref{eq:taun} and $\taucr = \taun(0)$ due to symmetry. The results are plotted for various periods $P$ of $\taup(x)$. (b) Error between the simulation results and the results obtain from the analytical approach. The error between the markers is calculated using the interpolation (dashed line in (a)).}
    \label{fig:SimulatonVsTheroy}
\end{figure}

In the previous section, we showed that coalescence of slip patches causes discontinuities in the growth of the largest slip patch. This may lead to a critical slip patch that becomes unstable even though shortly before, the most critical slip patch was much smaller than $\nuclen$. 
This coalescence-based process was not included in existing theory (see Sec.~\ref{sec:theory}), which considered a single localized monotonically evolving heterogeneity.
To verify the extent of this discrepancy, we compare here the simulation results with the theoretical prediction for a wide range of characteristic interface length $P/\nuclen$ and three different values of $\tauamp$.

We observe that the predictive power of the theoretical solution varies between the three nucleation regimes, as shown in Fig.~\ref{fig:SimulatonVsTheroy}. In the Uenishi \& Rice nucleation regime, \textit{i.e.}, for large $P/\nuclen$ where no pre-critical coalescence occurs (see Fig.~\ref{fig:DeterministicInterface}d), the prediction is generally good and improves even more for increasing $P/\nuclen$. Increasing discrepancies for $P/\nuclen \approx 1$ are likely related to second order effects due to the interaction between neighboring slip patches, which are in these configurations in particularly close proximity (when local maxima in $\taup$ are just at the border of $\nuclen$, \textit{e.g.}, Fig.~\ref{fig:DeterministicInterface}a). Note that the Uenishi \& Rice nucleation regime reaches $P/\nuclen < 1$ due to the superposition of $\hat \tau(x)$ onto the sinusoidal profile.

The discrepancy between theory and simulation reaches its maximum in the critical coalescence regime (see Fig.~\ref{fig:SimulatonVsTheroy}). This is due to very early and critical coalescence, which is not accounted for in the theoretical solution. The reason is that Eq.~\ref{eq:taun} computes the nucleation stress at a given point by integrating the $\taup(x)$ profile along $\pm \nuclen/2$. In this calculation, it does not know about the other local minima in $\taup(x)$ that are in close proximity but still further away than $\pm \nuclen/2$. Therefore, it does not know of the existence of the adjacent slip patch, caused by this near local minimum, and hence cannot predict coalescence. Consequently, the theoretical solution overestimates the strength of the interface at this point, which is confirmed by Fig.~\ref{fig:SimulatonVsTheroy}. 

In the sub-critical coalescence regime, the theoretical prediction works well again. 
For decreasing $P/\nuclen$, theory and simulation converge to $\hat{\tau}_\mathrm{cr}$, which is the critical nucleation stress for the $\hat \tau(x)$ profile (Eq.~\ref{eq:det_interface}). 
In this regime, coalescence happens within $\nuclen$, hence the theoretical solution accounts for the local minima in $\taup(x)$. Only the last coalescence, \textit{e.g.}, coalescence 5 in Fig.~\ref{fig:DeterministicInterface}f, is not included in the prediction, but its effect diminishes with decreasing $P/\nuclen$.

It is interesting to note, that the overall trend indicates an increase in macroscopic strength $\taucr$ for decreasing characteristic interface length, \textit{i.e.} $P/\nuclen$. This is consistent with observations by \cite{albertini_stochastic_2020} on stochastic interfaces. Here, in this deterministic configuration, the theoretical solution presents oscillations (see Fig.~\ref{fig:SimulatonVsTheroy}a). They are caused by the integration in Eq.~\ref{eq:taun} over length $\nuclen$ with varying fraction of period $P$ of the sinusoidal profile in $\taup(x)$. 

In summary, the comparison between theory and simulation shows that the theoretical prediction works generally well for deterministic interfaces except if the interface fails through nucleation by critical coalescence.

\subsection{Stochastic interfaces: sub-critical slip nucleation }\label{sec:results:stochastic}

\begin{figure}
    \centering
    \includegraphics[width=\figwthS]{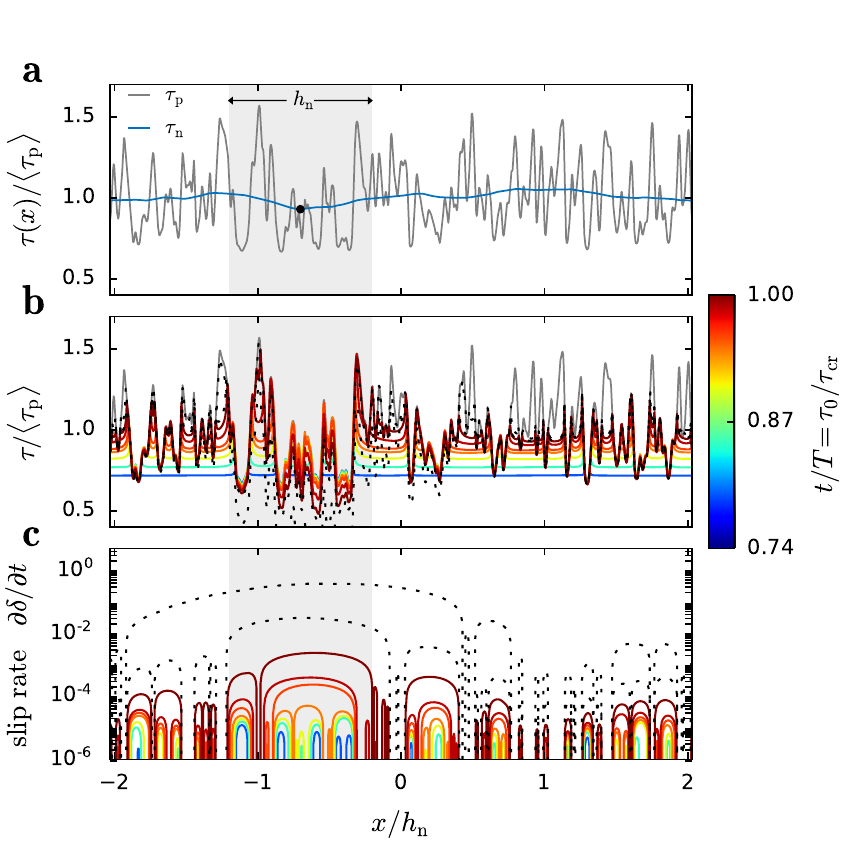}
    \caption{Coalescence of microslips in a random profile. 
    (a)~The random profile $\taup(x)$ with correlation length $\corlen/\nuclen=0.05$ shown as a gray line and the computed nucleation stress $\taun(x)$ depicted as a blue line. The black dot marks the minimum of $\taun(x)$ and corresponds to the location and stress level where global instability will nucleate. The gray area with width $\nuclen$ is centered around this critical point. 
    (b)~The random profile shown as a gray line and the stable stress state at various times is shown as colored lines with red being the last stable moment. Dashed lines show stress states during failure. The times $t$ are normalized by the time $T$ at which the profile fails and correspond to the load normalized by the interfacial strength $\tau_0/\taucr$ (color bar). 
    (c)~Slip rate along the profile marking the extension of the slip areas. The plotted time steps and colors match those in (b). The dashed lines show the dynamic slip propagation during failure.}
    \label{fig:RandomInterface}
\end{figure}

We now focus on interfaces with random strength profiles with relatively small correlation lengths, which have not been simulated by \cite{albertini_stochastic_2020}. 
First, we will consider a representative example shown in Fig.~\ref{fig:RandomInterface} to illustrate the nucleation process. In a second step, we will analyze the growth of slip patches for systems with various (small) correlation lengths. 

In these stochastic systems, the interface is characterized by a random profile of $\taup(x)$. Therefore, the location for nucleation of unstable slip is not obvious from $\taup(x)$. However, the profile of $\taun(x)$, computed through Eq.~\ref{eq:taun}, presents a minimum (see Fig.~\ref{fig:RandomInterface}a), which is a good prediction for the location of unstable slip nucleation \citep{albertini_stochastic_2020}.
Note that this point is not located at $\min \taup(x)$. The dynamic simulations confirm these observations (see Fig.~\ref{fig:RandomInterface}b and c).

We further observe that multiple sub-critical coalescence events occur before the interface transitions to macroscopic sliding.
The slip rate (see Fig.~\ref{fig:RandomInterface}c) shows the early formation of many small slip patches in areas of low strength. Because the random profile has randomly distributed local minima, the growth of existing patches, the formation of new ones and the coalescence of adjacent patches happen in a disordered manner in contrast to the deterministic cases presented in the previous section. 
However, the formation of large nucleation patches is mainly caused by a stepwise coalescence of several small patches and is not due to growth of large individual patches. Note that the coalescence events show a hierarchical pattern. Small patches coalesce first, then medium size patches coalesce to form a large patch. This is in contrast to the deterministic case, where the largest patch grew by constant increments at each coalescence event. Eventually, at $\tau_0=\taucr$ the size of the largest nucleation patch is close to the predicted $\nuclen$. It then becomes unstable and propagates dynamically through the whole system.

\begin{figure}
    \centering
    \includegraphics[width=\figwthS]{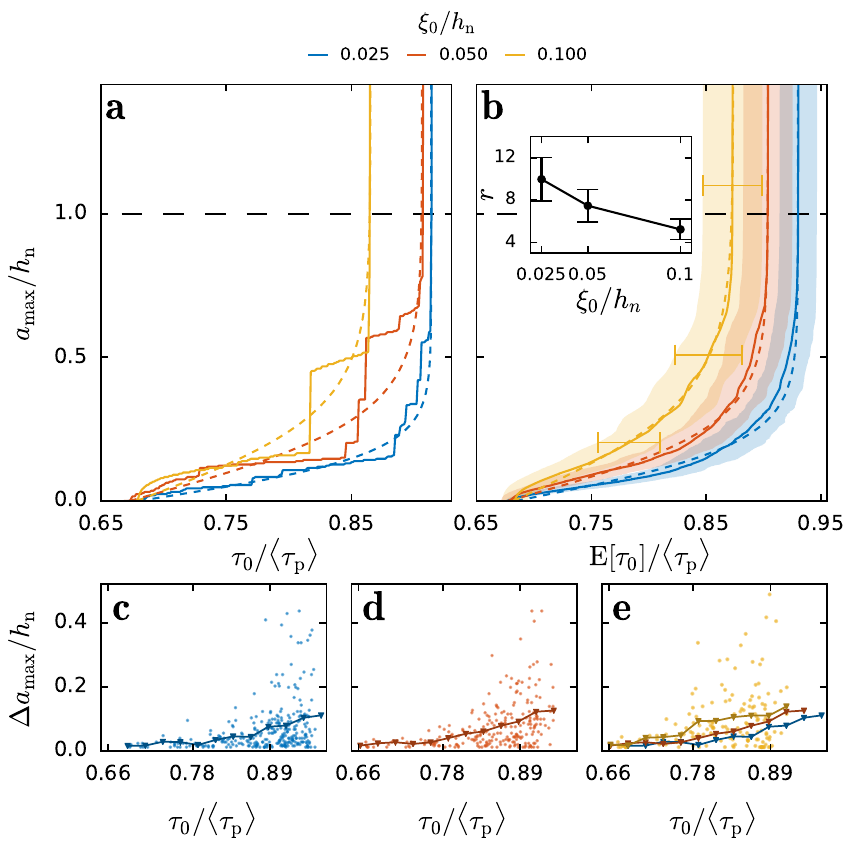}
    \caption{Slip patch growth. 
    (a)~Size of the largest slip patch $\patchsize$ with increasing load $\tau_0$ as a continuous line and logistic regression (Eq.~\ref{eq:logistic}) as a dashed line. For each $\corlen/\nuclen$ one representative example is shown. The $\corlen/\nuclen = 0.05$ example corresponds to the one shown in Fig.~\ref{fig:RandomInterface}.
    (b)~$\patchsize$ versus expected load $\mathrm{E}[\tau_0]$ based on 30 simulations for each $\corlen/\nuclen$ (continuous line) with one standard deviation error band. The logistic regression is depicted as a dashed line. The inset shows the average logistic rate $r$ for each $\corlen/\nuclen$ with a one standard deviation error bar.
    (c), (d) and (e)  Coalescence event sizes $\Delta \patchsize$ of the largest slip patch of all simulations for each $\corlen/\nuclen$, respectively. Data is reported before the onset of instability  \textit{i.e.}, $\patchsize(\tau_0) \leq \patchsize(\taucr)$ and for $\Delta \patchsize$ greater than twice the spatial discretization.
    The continuous line shows the average $\Delta \patchsize$ computed for bins of size $0.025\tau_0/\langle \taup \rangle$ (markers represent center of bin). All average $\Delta \patchsize$ are reported in (e) to allow for comparison.
    }
    \label{fig:RandomInterfaceStatistics}
\end{figure}

How does the critical slip patch grow during this process? We track the size of the largest slip patch $\patchsize$ (see Fig.~\ref{fig:RandomInterfaceStatistics}a) and observe that growth occurs smoothly for some periods but is regularly intermitted by small and large discontinuities. These jumps correspond to slip patch coalescence events. Note that the largest slip patch may not always remain the same patch because the coalescence of two smaller areas could exceed an existing larger patch. Generally, the growth of $\patchsize$ is initially rather slow but accelerates continuously. This accelerating effect is general as demonstrated by the expectation of the load $\mathrm{E}[\tau_0]$ required to cause a slip patch of size $\patchsize$ computed from 30 independent simulations (see Fig.~\ref{fig:RandomInterfaceStatistics}b). These observations are valid for simulations with various (generally small) correlation lengths, as shown for $\corlen/\nuclen=[0.025, 0.05, 0.1]$ in Fig.~\ref{fig:RandomInterfaceStatistics}b. Here, we further observe that systems with smaller correlation lengths require, on average, a smaller load for the largest slip patch to reach the critical length, and hence the macroscopic strength is lower. This confirms the results from the deterministic systems (see Sec.~\ref{sec:results:prediction}) and previous observations on stochastic systems by \cite{albertini_stochastic_2020}.

The acceleration in slip patch growth is directly linked to the coalescence size, \textit{i.e.}, the increase in size of the maximal slip patch caused by coalescence. This coalescence size $\Delta \patchsize$ varies significantly during the nucleation process (see Fig.~\ref{fig:RandomInterfaceStatistics}c-e). For all correlation lengths considered, we observe that there is a tendency for larger coalescence events as the loading proceeds. Nevertheless, small coalescence event also occur at later stages of the growth process. Hence, the variability of $\Delta \patchsize$ gradually increases with $\tau_0$, and $\Delta \patchsize$ can even be larger than $\corlen$ and exceed $0.4 \nuclen$.
Further, we observe that the growth of the mean $\Delta \patchsize$ as function of $\tau_0$ is higher for larger correlation length (see Fig.~\ref{fig:RandomInterfaceStatistics}e). This is consistent also with the following observations. The total number of coalescence events detected over 30 simulations are 295, 237 and 153 for $\corlen/\nuclen= 0.025, 0.05, \text{ and } 0.1$, respectively, and the inter-coalescence continuous slip patch growth represents, in average, 18\%, 27\%, and 43\% of the nucleation phase for $\corlen/\nuclen= 0.025, 0.05, \text{ and } 0.1$, respectively.
All in all, this shows that larger correlation lengths of the interface tend to support earlier and larger coalescence events, which results in less coalescence events over the course of the nucleation but faster slip patch growth and hence lower $\taucr$ since $\nuclen$ is reached more rapidly.

Can we provide an analytical expression for this growth process during nucleation? In view of $\nuclen$ being a constant, we propose to describe $\tau_0$ as function of $\patchsize$ since this will enable us, in a second step, to make a prediction for macroscopic friction strength. Based on the observed growth process in Fig.~\ref{fig:RandomInterfaceStatistics}b, we apply a logistic regression, which follows
\begin{equation}\label{eq:logistic}
    \tau_0(\patchsize) \approx \taup^{\min} + \left(\taucr^{\mathrm{ext}}-\taup^{\min}\right)\left[\frac{2}{1+e^{-r \,\patchsize/\nuclen}} -1\right] 
\end{equation}
with two fitting parameters, which are the logistic rate, $r$, and the stress $\taucr^{\mathrm{ext}}$. When ${\patchsize \rightarrow \nuclen}$, the term within the square-brackets goes to $1$, and, we find ${\tau_0(\patchsize \rightarrow \nuclen) = \taucr^\mathrm{ext}}$. Therefore, $\taucr^\mathrm{ext}$ corresponds the critical strength. In a later step, we will apply this regression on a smaller range of patch sizes $\patchsize<\nuclen$, and hence, $\taucr^\mathrm{ext}$ will become an extrapolated prediction of the critical strength.


The logistic regression works relatively well to describe single simulations (see dashed lines in Fig.~\ref{fig:RandomInterfaceStatistics}a). However, large discrepancies appear whenever large coalescence events occur (\textit{e.g.} at $\tau_0/\langle \taup \rangle \approx 0.82$ for simulation with $\corlen/\nuclen=0.1$ in Fig.~\ref{fig:RandomInterfaceStatistics}a). These discrepancies disappear when we apply the logistic regression on the expectation $\mathrm{E}[\tau_0]$ (see Fig.~\ref{fig:RandomInterfaceStatistics}b).
We observe that the logistic rate $r$ is smaller for larger $\corlen$, which results in lower $\mathrm{E}[\tau_0]$ for a given $\patchsize/\nuclen$.

\begin{figure}
    \centering
    \includegraphics[width=\figwthS]{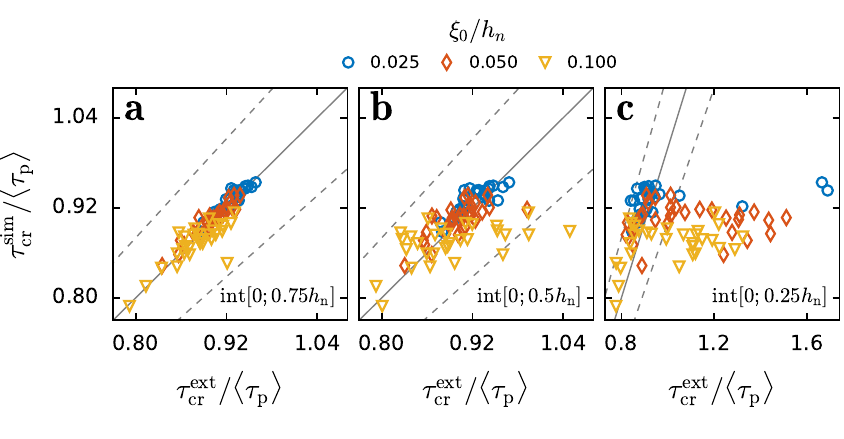}
    \caption{Prediction of the critical load. (a) Critical load of the simulation $\tau_\mathrm{cr}^\mathrm{sim}$ plotted against the extrapolated critical load $\tau_\mathrm{cr}^\mathrm{ext}$ using a logistic regression  (Eq.~\ref{eq:logistic}) interpolated over different intervals.  (a) interpolates the $\patchsize$ in the interval $[0;0.75\nuclen]$. (b) and (c) interpolate over an earlier nucleation state of $\patchsize<0.5\nuclen$ and $\patchsize<0.25\nuclen$, respectively. The 10\% error range is depicted as dashed lines. 
    }
    \label{fig:extrapolation}
\end{figure}

Finally, we test if the logistic regression can be used to predict macroscopic strength by fitting it to sub-periods of slip patch growth. 
The fitting parameter of interest is the extrapolated critical stress $\taucr^\mathrm{ext}$, which is then compared with the actual critical stress from the simulation $\taucr^\mathrm{sim}$. For relatively long regression windows ($\patchsize \in [0;0.75\nuclen]$), the prediction works quantitatively well as shown in Fig.~\ref{fig:extrapolation}a. However, for shorter regression windows, which would be more useful for prediction purposes, the precision decreases (see Fig.~\ref{fig:extrapolation}b and c).
Interestingly, the proposed extrapolation procedure performs better for cases with small correlation length because large coalescence events $\Delta \patchsize$ are less likely, especially in the early and mid-nucleation phase.

\section{Discussion}
\label{sec:discussion}

\subsection{Physical model}

In the current study we considered the nucleation of frictional ruptures along interfaces with spatially heterogeneous friction strength embedded in a two dimensional linear elastic periodic  medium. Moreover, we assumed a linear slip-weakening friction law and uniform loading.
These assumptions allowed us to study the nucleation problem efficiently using dynamic simulations and analytical theory.
However, real systems are three dimensional and friction laws can depend on the age of contact \citep{dieterich_direct_1994,dillavou_nonmonotonic_2018} and on the rate of sliding \citep{rabinowicz_intrinsic_1958,kilgore_velocity_1993,baumberger_solid_2006}. Additionally, in finite-size systems, the applied loading on the interface is non-uniform due to geometry.

We assumed a linear-slip weakening friction law because it is the simplest law that can represent the experimentally observed dynamic weakening during dynamic frictional rupture propagation \citep{svetlizky_dynamic_2020,kammer_fracture_2019}. 
Rate-and-state dependent friction laws, as introduced by \cite{dieterich_modeling_1979} and \cite{ruina_slip_1983}, would also capture the dynamic weakening and generally represent well the experimentally observed rate dependence of friction at velocities typical of the nucleation regime. However, the concept of static friction coefficient is not clearly represented by the most common rate-and-state laws because the formulations usually diverge for zero slip rate.
Nucleation on rate-and-state interfaces has been studied  by \cite{ray_earthquake_2017,ray_homogenization_2019} using stability analysis. Considering nucleation on random interfaces with rate-and-state friction will be addressed in future work. 

We considered contact between thin plates and, therefore, assumed a two dimensional medium. This assumption is valid for cases where the critical nucleation length is considerably larger than the plate thickness, such that three dimensional effects only occur during the initial phase of the nucleation, when the slip patch size is smaller than the plate thickness. 
However, if the nucleation length is similar or even smaller than the plate thickness, we expect three dimensional effects to play an important role.
An interesting mechanism has been observed on brittle ruptures in three dimensional media, where the stress intensity factor is affected by the local curvature of the front \citep{rice_first-order_1985,lebihain_effective_2020}. Thus, the characteristic length of the interface properties would directly affect the effective front strength due to its non-planar geometry.

\subsection{Interpretation of nucleation patterns}
\label{sec:disc:deterministic}

By considering deterministic frictional interfaces, we were able to create a controlled nucleation process and to study different formation patterns of critical slip patches.  We showed that initiation of slip varies depending on the characteristic length of the interface and that it can be categorized into three regimes. 

Large correlation lengths correspond to the "Uenishi \& Rice nucleation" (UR nucleation), which is characterized by the continuous growth of non-interacting slip patches.
This nucleation process resembles experimental observations \citep{nielsen_experimental_2010, latour_characterization_2013, fukuyama_spatiotemporal_2018}, where only a single rupture front was recorded.
Further, for very large correlation lengths the system approaches the homogeneous limit.

We defined the regime of "critical coalescence" when a critical nucleation patch is suddenly formed by coalescence of nucleation patches considerably smaller than the critical nucleation length. Such a behavior was observed by \cite{mclaskey_foreshocks_2013} in some of their experiments. They recorded the formation of multiple ruptures and observed that the onset of global instability was initiated by the coalescence of two rupture fronts.

If the critical nucleation patch grows by a stepwise coalescence of its neighboring patches, we classified it as "sub-critical coalescence". To our knowledge, there has been no experimental evidence supporting the existence of "sub-critical coalescence". However, it may not have been detected yet because of lacking spatial resolution in experimental measurements. A sub-critical coalescence process could appear as continuous nucleation, \textit{i.e.}, Fig.~\ref{fig:Patchgrowth}d would become a continuous line.
The case of very small correlation length corresponds to the homogenization limit, where the system's response corresponds to its average properties.

In Fig.~\ref{fig:SimulatonVsTheroy} we compared the critical stress of the simulation with the analytical approach introduced by \cite{uenishi_universal_2003}. We noted large discrepancies in the regime of "critical coalescence" and for the UR nucleation regime with small $\corlen/\nuclen$ (Fig.~\ref{fig:SimulatonVsTheroy}b). Intuitively, we would expect large errors for critical coalescence only. 
However, for the UR nucleation with small $\corlen/\nuclen$, the presence of closely spaced interacting slipping regions affects the nucleation process. The theoretical model does not account for this effect and hence overestimates the strength.

\subsection{Interpretation of stochastic interfaces}
\label{sec:disc:stochastic}
With this study, we not only showed the mechanism of slip nucleation by coalescence of microslip on deterministic interfaces but also verified it on random frictional interfaces with small correlation lengths compared to the critical nucleation length. We showed that for a given slip patch size the expected (average) load value follows a logistic function with increasing logistic rate for smaller correlation length. While the logistic regression describes well the expectation curves (Fig.~\ref{fig:RandomInterfaceStatistics}b), it  has limitations for predicting individual examples because of large coalescence events leading to sudden large increments followed by a plateau (see case $\corlen/\nuclen=0.1$ in Fig.~\ref{fig:RandomInterfaceStatistics}a). 
The coalescence size is a random variable with non-stationary properties as function of applied load. We have shown that large coalescence events occur earlier in the nucleation phase when the correlation length is large (see Fig.~\ref{fig:RandomInterfaceStatistics}c-e), leading to lower interface strength and higher strength variation.
Interestingly, we also noted that phases of continuous propagation are longer for larger correlation length, as well as the cumulative continuous propagation.
In these disordered systems, the growth of the largest slip patch is analogous to a random walk with varying step size. The coalescence size corresponds to the random step taken at each load increment. With increased loading, large coalescence events become more likely.

The nucleation of in-plane frictional ruptures has been investigated by
\cite{latour_characterization_2013} in laboratory experiments. By measuring the evolution of the rupture length as a function of time, they identified three phases during nucleation: A quasi-static, accelerated and dynamic propagation phase. They observed an exponential growth during the quasi-static phase, and an inverse power law in the acceleration phase.
However, they report a single nucleation patch in their experiments, whereas in our simulation many small patches nucleate and coalesce to a large one. Consequently, the laboratory situation would rather correspond to a very large correlation length with a single slip patch that grows in size continuously. 
However, distinct spatial patterns are visible in their measurements \citep[fig.~1]{latour_characterization_2013}, which may represent interface heterogeneity. As shown in Fig.~\ref{fig:RandomInterface}c, the maximum slip patch has larger slip rates compared to smaller less critical patches. This implies that small sub-critical patches may be harder to visualize in experiments.
Further, the average curves in Fig.~\ref{fig:RandomInterfaceStatistics}b show a similar behavior as the laboratory observations: A quasi-static crack growth that passes into a acceleration phase before the crack starts propagating dynamically. 
The transition from the quasi-static to the acceleration phase corresponds to an increased probability for larger coalescence events (see Fig.~\ref{fig:RandomInterfaceStatistics}c-e). However, due to the discrete nature of the increase in patch size, in some cases the acceleration phase is characterized by a large coalescence followed by a plateau (see $\corlen/\nuclen=0.1$ in Fig.~\ref{fig:RandomInterfaceStatistics}a).

Further, we used the logistic model to predict the strength of the interface from its early propagation before the onset of instability (see Fig.~\ref{fig:extrapolation}). The prediction is reliable if the growth of $\patchsize$ is well advanced.
Interestingly, the prediction for an interpolation interval of 50\% the critical nucleation length yields most results within 10\% error. However, for smaller intervals the prediction is unreliable because the system did not enter the acceleration phase yet.
For very small correlation lengths ($\corlen/\nuclen=0.025$), the prediction works reasonably well, even for a small interpolation range. However, few outliers deviate significantly more than they do for interfaces with larger correlation length. This approach provides an empirical equation of motion for the nucleation of slip patches. It can be useful to analyze the stability and strength of frictional interfaces before failure has occurred. 

\cite{albertini_stochastic_2020} reported discrepancies between simulations and the theoretical model for relatively small correlation lengths (${\corlen/\nuclen \lesssim 0.25}$). 
Unlike for the deterministic system, we cannot strictly distinguish between the three nucleation patterns, as observed in Sec.~\ref{subsubsec:NucleationPattern}. However, the observed trends are comparable. For larger correlation lengths, the nucleation process is most likely following a UR nucleation because coalescence is less likely. Notably large coalescence events leading to nucleation by critical coalescence are not likely to occur and hence the analytical solution by \cite{uenishi_universal_2003} is expected to work generally well. However, for correlation lengths in the intermediate range (the small values in \citep{albertini_stochastic_2020}), the probability for coalescence increases and, therefore, it is more likely for critical coalescence to cause discrepancies compared to the theoretical prediction. Finally, for small correlation lengths, coalescence is ubiquitous and sub-critical coalescence is likely to be the dominant slip patch growth pattern.

\section{Conclusion}
\label{sec:conclusion}

We studied the nucleation of slip patches on linear slip-weakening interfaces with heterogeneous friction properties. We considered a uniform loading and, both, deterministic and stochastic non-uniform local friction properties.
In the deterministic setup, we systematically varied the characteristic length of the interface strength profile and observed three distinct nucleation patterns: smooth growth of a slip patch as observed by \cite{uenishi_universal_2003}, growth by critical coalescence, and sub-critical coalescence.

A smooth growth of the critical slip patch is observed when the characteristic length is large compared to the critical nucleation length. In this case the global interface strength is well represented by an existing analytical model. 
For intermediate characteristic lengths, a sudden increase in the size of the nucleation patch to a size significantly larger than the critical nucleation length was caused by the coalescence of two nucleation patches considerably smaller than the critical nucleation length (critical coalescence). Since the analytical approach does not account for large coalescence, the actual global strength is overestimated.
For characteristic lengths considerably smaller than the critical nucleation length, a series of coalescence events causes a stepwise increase of the nucleation patch (sub-critical coalescence), which results in a precise prediction by the analytical model.

We generalized our study by considering a stochastic interface and showed that for small correlation length the nucleation mechanism is similar to the sub-critical coalescence in the deterministic case.  
However, the size of each coalescence is a non-stationary random variable with respect to the load and can exceed the correlation length. Hence, the nucleation process progresses in a disordered manner with large coalescence events being more likely in the late nucleation phase.
Therefore, we provide an empirical formula to describe the sub-critical nucleation process: The expectation of the load as function of the largest slip patch size follows a logistic function. 
This simple equation relates the loading stress with the size of the largest nucleation patch and allows us to extrapolate for the critical nucleation length. As a result, the global interface strength can be estimated before failure has occurred.



\printcredits

\bibliographystyle{cas-model2-names}

\bibliography{references_dsk,references_ss,references_gab}

\newpage\null\thispagestyle{empty}\newpage

\end{document}